\documentclass[twocolumn,showpacs,pre]{revtex4}
\usepackage{amsmath,amssymb,amsfonts,amsthm}
\usepackage{graphicx}

\begin{document}

\newcommand{\be}{\begin{equation}}
\newcommand{\ee}{\end{equation}}
\newcommand{\bs}[1]{\begin{equation}\label{#1}\begin{split}}
\newcommand{\es}{\end{equation}}
\newcommand{\ci}{\vec{c}_i}
\newcommand{\Gi}{{{\cal G}_i}}
\newcommand{\Li}{{{\cal L}_i}}
\newcommand{\x}{\vec{x}}
\newcommand{\bsy}{\boldsymbol}
\renewcommand{\r}{\vec{r}}
\renewcommand{\u}{\vec{u}}
\renewcommand{\vec}{\mathbf}
\newcommand*\rfrac[2]{{}^{#1}\!/_{#2}}
\newcommand{\ufrac}[2]{{\rm #1}/{\rm #2}}
\newcommand{\unit}[1]{\,\mbox{#1}}

\title{Flow force and torque on submerged bodies in lattice-Boltzmann via momentum
exchange}
\author{Juan P. \surname{Giovacchini}$^{1,3}$}
\email{giovacchini@famaf.unc.edu.ar}
\author{Omar E. \surname{Ortiz}$^{2,3}$}
\email{ortiz@famaf.unc.edu.ar}

\affiliation{$^1$Departamento de Mec\'anica Aeron\'autica, Instituto
Universitario Aeron\'autico, C\'ordoba, Argentina.\\
$^2$Facultad de Matem\'atica, Astronom\'{\i}a y F\'{\i}sica, Universidad
Nacional de C\'ordoba, Argentina.\\
$^3$Instituto de F\'{\i}sica Enrique Gaviola (CONICET), C\'ordoba,
Argentina.}

\begin{abstract}
We present a new derivation of the momentum exchange method to compute the flow
force and torque on a submerged body in lattice Boltzmann methods. Our
derivation does not depend on a particular implementation of the boundary
conditions at the body surface and relies on general principles. We recover some
well known expressions, in some cases with slight corrections, to treat the
cases of static and moving bodies. We also present some numerical tests that
support the correctness of the formulas derived.
\end{abstract}

\pacs{47.11.-j, 47.10.-g, 51.10.+y}

\maketitle

\section{Introduction}

During the last twenty-five years the Lattice-Boltzmann methods (\emph{LBM})
have been greatly developed in many aspects. Today they can be used, to treat
multiple problems involving both compressible and incompressible flows on simple
and complex geometrical settings.

It is of crucial importance, in many applications that involve moving bodies
surrounded by a fluid flow, to have a good method or algorithm to compute the
flow force and torque acting on the bodies. By good we mean a method that is
simple to apply, that is accurate and fast, so as not to spoil the efficiency of
the flow computing method.

The classical way to compute forces, and so torque, on submerged bodies is via
the computation and integration of the stress tensor on the surface of the body. In 
LBM the stress tensor is a local variable, its computation and extrapolation 
from the lattice to the surface is computationally expensive, which ruins the
efficiency of the LBM. However, this method is widely used in LBM 
\cite{Inamuro:2000, Xia:2009, Li:2004}.

In 1994 Ladd introduced a new method, the \emph{momentum exchange} (\emph{ME}), to
compute the flow force on a submerged body \cite{Ladd:1994, Ladd:1994b}. 
Ladd's idea was rather heuristic and very successful, where the force is obtained by
accounting the exchange of momentum between the surface of the body and the fluid,
the latter being represented by ``fluid particles'' whose momentum is easily
written in terms of the LBM variables that describe the fluid at the mesoscopic scale.
Aidun et. al. \cite{Aidun:1998} introduce some improvements to Ladd proposal, 
obtaining a robust method to analyze suspended solid particles, and excluding the 
simulation of the interior fluid with a modified midway bounce-back boundary 
condition.
Then, using boundary condition method to arbitrary geometries, Mei et. al. 
\cite{Mei:2002_pre65-041203} proposed a method to evaluate the fluid forces
from the idea of ME.

The ME algorithm is specifically designed and adapted to LBM; it is therefore
more efficient than stress integration from the computational point of view.

The ME algorithm has been tested and applied successfully to a variety of
problems \cite{Ladd:1994b, Mei:2002_pre65-041203, Lallemand:2003}. 
For the mentioned ME methods, except the presented in \cite{Aidun:1998}, 
some accuracy problems have been detected though, when applied to moving bodies 
\cite{Li:2004, Wen:2012}. 

Some approaches to improve the methods in problems with 
moving bodies were made. Wen et. al. \cite{Wen:2012}, based in the proposal  
of \cite{Aidun:1998} gives corrections terms to the forces given from 
\cite{Mei:2002_pre65-041203}. Others alternative improved ME methods, based 
in the evaluation of force respect to a moving frame of reference, were 
proposed in \cite{Wen:2014}.

The main goal of this paper is to provide a formal derivation of
the momentum exchange algorithm. This new derivation provides in turn, some
corrections to the Mei et. al. \cite{Mei:2002_pre65-041203} formula and also to 
some newer, improved versions of momentum exchange algorithm that have been 
proposed \cite{Wen:2012,Wen:2014}.

The rest of the paper is organized as follows. In section \ref{sec_lbm} we
briefly discuss the lattice-Boltzmann method with the main purpose of
introducing notation; the method used to treat boundary conditions is also
explained in this section. In Section \ref{sec:momentum_exchange}, the core of
the paper, we present a derivation of the momentum exchange method to
determine both, the flow force and torque on static or moving bodies. In section
\ref{sec:numerical_tests} we present two numerical tests where we implement
the methods derived in section \ref{sec:momentum_exchange}.  
In section \ref{sec:comments} we make some comments.

\section{The lattice-Boltzmann method}\label{sec_lbm}

In this section we present the basic equations of the lattice Boltzmann methods
with the main purpose of introducing the notation used along the paper. For a
thorough description of the Boltzmann equation we refer to
\cite{Harris:2004,Sone:2007}. For a more complete presentation of LBM we refer to
\cite{He:1997,Succi:2001,Wolf-Gladrow:2000}.

The Boltzmann equation (\emph{BE}) governs the time evolution of the
single-particle distribution function $f(\x,\bsy{\xi},t),$ where $\x$ and
$\bsy{\xi}$ are the position and velocity in phase space. The lattice Boltzmann
equation (\emph{LBE}) is a discretized version of the Boltzmann equation, where
$\x$ takes values on a uniform grid (the lattice), and $\bsy{\xi}$ is not only
discretized, but also restricted small number of values \cite{Luo:1997}. By far
the models used most frequently are the ones with collision integral simplified
according to the Bhatnagar, Gross, and Krook (\emph{BGK}) approximation
\cite{Bhatnagar:1954} with relaxation time $\tau$. In an
isothermal situation and in the absence of external forces, like gravity,  the
LBE of this models read
\begin{multline}\label{eq:LBE}
f_i(\x_A+\ci\delta t, t+\delta t) = f_i(\x_A, t) -
\frac{1}{\tau}\Bigl(f_i(\x_A, t)\\
- f^{eq}_i(\x_A, \rho, \u, t)\Bigr),\\
i=0,1,\dots,Q-1.
\end{multline}
Here $f_i=\omega_i f(\x_A,\ci,t)$ is the $i$-th component of the discretized
distribution function at the lattice site $x_A,$ time $t,$ and corresponding to
the discrete velocity $\ci$. $\omega_i$ is the
$i$-th quadrature weight (explained below), and $Q$ the number of discrete
velocities in the model.  In compressible-flow models the lattice constant
$\delta x,$ that separate two nearest neighbor nodes, and the time step $\delta
t$ are related with the speed of sound $c/\sqrt{3}$ by $\delta x = c \delta
t$ \footnote{In incompressible-flow models, the same relation between $\delta x$
and $\delta t$ holds, but the constant $c$ is no longer related to the speed of
sound.}. The coordinates of a lattice node are $\x_A$, where the integer multi
index $A=(j,k,l)$ (or, $A=(j,k)$ in the two-dimensional case) denotes a
particular site in the lattice. The equilibrium distribution function $f^{eq}$
is a truncated Taylor expansion of the Maxwell-Boltzmann distribution. It is
this approximation one of the reasons that makes LBM accurate only at low Mach 
numbers \cite{Luo:1997}.

The macroscopic quantities such as the fluid mass density $\rho(\x,t),$ and velocity
$\u(\x,t)$, are obtained, in Boltzmann theory, as marginal distributions of $f$
and $\bsy{\xi} f$ when integrating over $\bsy{\xi}$. In LBM this integrals are
approximated by proper quadratures. Specific values of $c_i$'s and $\omega_i$'s,
$i=0,1,\dots,Q-1,$ are made so that these quadratures give exact results for the
$\bsy{\xi}$-moments of order 0, 1 and 2 \cite{Luo:1997,Wolf-Gladrow:2000}. We
have
\begin{equation}\label{eq:LBMmacroscopics_0}
\rho(\x_A,t) = \sum_{i=0}^{Q-1}f_i(\x_A,t),
\end{equation}
and
\begin{equation}\label{eq:LBMmacroscopics_1}
\rho \u(\x_A,t) = \sum_{i=0}^{Q-1} \ci f_i(\x_A,t).
\end{equation}
In the simulations we present in this paper, we are interested in incompressible
flow problems, where we modify Eq. \ref{eq:LBMmacroscopics_1} according to the
quasi-incompressible approximation presented in \cite{He:1997b}. In this
approximation $\rho$ is replaced by $\rho_0$, a constant fluid mass density.
 
A single time step of the discrete evolution equation \eqref{eq:LBE} is frequently 
written as a two-stage process
\begin{multline}
\hat f_i(\x_A, t) = f_i(\x_A, t) - \frac{1}{\tau}\Bigl(f_i(\x_A, t) \\
- f^{eq}_i(\x_A,\rho,\u,t)\Bigr), \label{eq:collision}
\end{multline}
and
\begin{equation}
f_i(\x_A+\ci\delta t, t+\delta t) = \hat f_i(\x_A, t).\label{eq:streaming}
\end{equation}
The computation of $\hat f_i$ on the whole lattice, Eq. \eqref{eq:collision}, is
called the \emph{collision step}, while the computation of $f_i$ at $t+\delta
t$, Eq. \eqref{eq:streaming}, on the whole lattice is called \emph{streaming
step.}

\subsection{Treatment of boundary conditions}\label{subsec:BoundaryConditions}

Many methods have been proposed in the literature to implement boundary
conditions on moving boundaries with complex geometries in LBM. The method
introduced in \cite{Filippova:1998}, later improved in
\cite{MeiLuoShyy:1999,MeiShyyYuLuo:2000}, has been extensively tested and is the
one we use in the simulations presented in this paper\footnote{The first method
proposed to impose boundary conditions in LBM is known as bounce-back.
Bounce-back was appropriate to treat rectilinear boundaries which are aligned
with the lattice. The application of bounce-back when the boundary is of general
shape would be equivalent to approximate the boundary of the body by a
stair-step shape boundary coincident with lattice links, which implies a loose
of accuracy.}. We explain this method briefly in what follows.  We emphasize
that our derivation of momentum exchange is completely independent of the
boundary condition method selected to perform the numerical tests.

\begin{figure}[t]
\begin{center}
\includegraphics[scale=0.35]{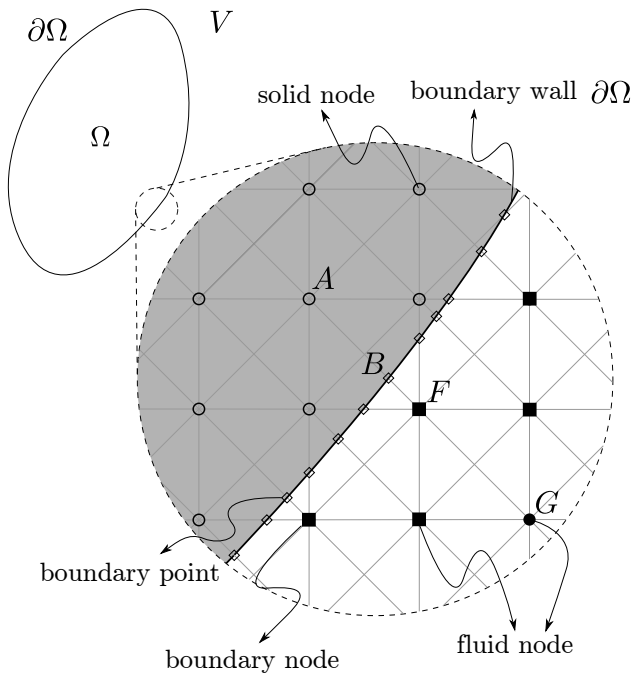}
\caption{Detail of boundary region, surrounding fluid and lattice.}
\label{fig_01}
\end{center}
\end{figure}

We consider a body that fills a region $\Omega$ with closed boundary
$\partial\Omega$ immersed in a fluid flow, and concentrate in a small portion of
the boundary and its surrounding fluid as shown in Figure \ref{fig_01}. The
lattice nodes and links are also shown in the figure. Empty circles represent
nodes lying inside the body region (solid nodes), while filled circles and
squares represent nodes lying in the fluid region at the time shown.
At time $t$ a piece of boundary lie, in general, between lattice nodes.
Consider a node $F$ on the fluid with a neighbour node $A$ inside the body.  To
determine the values of $f_i(\x_F=\x_A+\ci\delta t, t+\delta t)$, the streaming
step needs ``non-existent'' information coming from node $A$. It is the
LBM implementation of the boundary conditions what provides this information with the
desired accuracy.

The implementation of boundary conditions in LBM can be thought, at mesoscopic
scale, as the introduction of a fluid flow inside $\Omega.$ It is this
artifficial flow what provides the needed information to evolve the outer flow
so that it satisfies the right macroscopic boundary conditions at
$\partial\Omega.$ Even when the boundary $\partial\Omega$ is a physical boundary
for the fluid, the mesoscopic LBM description of the fluid allow the fluid
``particles'' to stream across the surface $\partial\Omega$, both from inside
out and viceversa.

We present here some particular proposals that will be used in section
\ref{sec:numerical_tests}. From now on we refer as ``boundary nodes'' those
lattice nodes on the fluid side, like $F,$ that are involved in the imposition
of boundary conditions.

The method presented in \cite{Filippova:1998} proposes to determine
$\hat f_i(\x_A,t)$ so that the linearly interpolated velocity at the boundary
point $B$ is the correct boundary velocity at that point. This is
\begin{multline}\label{eq:Fillipova_interpolation}
\hat f_i(\x_A, t) = (1-\chi)\hat f_{\bar{i}}(\x_A+\ci\delta t) + \\
\chi g_{\bar{i}}(\x_A, t) + 
2\omega_{\bar{i}}\rho\frac{3}{c^{2}} \vec c_{\bar{i}}\cdot\u_B
\end{multline}
where $\bar i$ denotes the index for
the opposite direction to $\vec{c}_i$ (i.e., $\vec{c}_{\bar i} = -\vec{c}_i$),
and
\begin{multline}\label{eq:Fillipova_equilibrium}
g_{\bar{i}}(\x_A, t) = \omega_{\bar{i}}\rho(\x_A+\ci\delta t) \Bigl(
1+\frac{3}{c^{2}}\vec c_{\bar{i}}\cdot\u_{bf}+\\
\frac{9}{2c^{4}}(\vec c_{\bar{i}}\cdot\u_F)^{2}-
\frac{3}{2c^{2}}\u_F\cdot\u_F \Bigr)
\end{multline}
is a fictitious equilibrium distribution function at the fluid node $A.$
$\omega_i,~i=0,1,\dots,Q-1,$ are the weight factors of the LBM method. 
$\u_B=\u(\x_B,t)$ and $\u_F=\u(\x_F,t)$ are the boundary and fluid velocities 
respectively, with $\x_B$ the intersection point between the boundary and the 
link joining $A$ with $F.$ 
Different choices of $\u_{bf}$, a velocity between $\u_B$ and $\u_F$, give 
alternative values of the parameter $\chi$, the weighting factor
that controls the interpolation (or extrapolation). To improve numerical stability
\cite{MeiLuoShyy:1999,MeiShyyYuLuo:2000} propose
\[
\u_{bf}=\u_{G}=\u(\x_F+\ci\delta t,t), \quad
\chi=\frac{2\Delta-1}{\tau-2}, \quad \text{if} \, \Delta<\frac{1}{2},
\]
and 
\[
\u_{bf}=\u_F + \frac{3}{2\Delta}(\u_B-\u_F), \quad
\chi=\frac{2\Delta-1}{\tau+\frac{1}{2}}, \quad \text{if} \,
\Delta\geq\frac{1}{2},
\]
where $0\leq\Delta\leq1$ is the fractional distance
\begin{equation}
\Delta  = \frac{\|\x_F-\x_B\|}{\|\x_F-\x_A\|}.
\end{equation}

When the body moves with respect to the lattice, there may be nodes in the body region
at time $t$ that become fluid nodes at time $t+\delta t$. It is then necessary
to assign initial values to the variables at the new fluid nodes to evolve them. A
practical way to do this is to evolve the nodes in the body region (solid nodes)
so that they have values assigned when they become fluid nodes. There are more
precise initializations for the variables at these nodes that change domain,
like the one proposed in \cite{Lallemand:2003}.

\subsection{Forces evaluation in lattice Boltzmann method}

It is of great interest to have a robust and accurate method to compute flow
forces in fluid mechanics. Several algorithms have been proposed to carry out
this in the context of LBM.  Many of these procedures fall in one of the
categories: stress integration (\emph{SI}) or momentum exchange (\emph{ME}).
Stress integration is based on the classical hydrodynamic approach (see e.g.,
\cite{Inamuro:2000}). In the context of LBM, the computational performance of ME
is higher than that of SI. In SI one needs to compute the stress tensor in all
lattice nodes which are near neighbors of the body surface. One then needs to
extrapolate the stress tensor to the surface, and finally obtain the total flow
forces on the body as an integral over the whole body surface. In ME the
procedure is simpler. The total force on the body is the sum of all
contributions due to momentum change, in the directions pointing towards the
body surface, over all boundary nodes.

In this section we write forces in general when we mean either force or torque.
The idea of forces evaluation via momentum exchange was introduced by Ladd
\cite{Ladd:1994,Ladd:1994b} as a heuristic algorithm by thinking the flow as
composed by ``fluid particles'' and using particle dynamics to describe their
interaction with the boundaries.  In this method, a particle suspension model is
proposed where the same boundary condition procedure is applied for both
interior and exterior fluid, using in all cases a midway bounce-back boundary
condition. The forces evaluations are carried out considering the interior and
exterior fluid.

Based in the works of Ladd, Aidun et. al. \cite{Aidun:1998} introduce some 
improvements to Ladd proposal, obtaining a robust method to analyze suspended
solid particles with any solid-to-fluid density ratio. 
They also proposed a modified midway bounce-back as boundary condition, and exclude 
the simulation of the interior fluid. The forces are evaluated considering the exterior 
fluid plus an impulsive contribution due to the nodes that are covered or uncovered 
when the body of interest move inside the fluid.

Then, from the idea of momentum exchange, Mei et. al.
\cite{Mei:2002_pre65-041203} proposed a method to evaluate the fluid forces
acting on a submerged body using a boundary condition method applied to
arbitrary geometries.  They exclude the simulation of the interior fluid as done
in \cite{Aidun:1998}.  The direct application of this method to problems with
moving bodies fails to obtain accurate forces evaluation as was shown in
\cite{Li:2004,Wen:2012}.  Some proposals to improve the method presented
in \cite{Mei:2002_pre65-041203} for problems with moving bodies were made. Wen
et. al. \cite{Wen:2012} presented one of this proposals. Their correction is
based in the introduction of terms representing impulsive forces. Aidun et. al.
\cite{Aidun:1998} give an improved an accurate method in moving geometry
problems.
  
The impulsive force terms introduced in \cite{Aidun:1998} and \cite{Wen:2012},
come from the nodes that are covered or uncovered when the body moves 
with respect to the lattice. This correction provoked some controversies, the
main discussion being about some ``noise'' that appear in the evaluation of
forces.

Based on the work of Mei et. al. \cite{Mei:2002_pre65-041203}, other approaches
to evaluate forces in moving geometries, without the introduction of impulsive
terms, were made. No rigorous proof was presented for these methods.
Both \cite{Wen:2014} and \cite{Krithivasan:2014} present a similar methods that
are based in computing the momentum exchange in a reference frame comoving with
the wall.  

All the ME based methods cited here were specifically designed for LBM and have
been implemented and tested in many fluid-mechanical problems. To the knowledge
of the authors there is no formal derivation of them in the literature. The work
of Caiazzo and Junk present an analysis of ME that uses an asymptotic expansion
\cite{Caiazzo:2008}.

In this work we give a demonstration of ME, from a fluid mechanics perspective,
in which some terms previously introduced as ad-hoc corrections appear
naturally. In particular, we find that the corrections proposed in
\cite{Wen:2012} and \cite{Aidun:1998} are adequate when evaluating the force in
a reference frame fixed to the lattice. In the spirit of our deduction of ME, we
also deduce the alternative description presented in
\cite{Wen:2014,Krithivasan:2014}, which is based on a reference frame comoving
with the body.  

\section{Momentum exchange method}\label{sec:momentum_exchange}

We want to simulate a fluid flow around a submerged body, within a region of
space that we denote by $V.$ We consider $V$ to be a fixed region of space as 
seen on an inertial reference frame. We have covered $V$ with a uniform constant 
lattice to solve the fluid motion by applying the lattice-Boltzmann method as 
described in section \ref{sec_lbm}.

The submerged body occupies a sub-region $\Omega(t)\subset V$ that we consider,
along the whole simulation, strictly contained in $V$. As the time dependence
indicates, $\Omega(t)$ doesn't need to be fixed. $\Omega(t)$ can move and could
even change shape.
  
In this section we derive the force and torque that the flow applies on the body. 
The movement of the body is assumed to be prescribed along this derivation, i.e.,
$\Omega$ is a given function of $t$. During an actual computation the body
movement is determined by integrating the equations of motion of the body
simultaneously with the flow equations. The equations of motion of the body take
into account the fluid force on the body, the bulk forces like weight, etc.

\subsection{Reynolds transport theorem}

For future reference we briefly remind here the Reynolds transport theorem. We consider first the
case of a fluid system. Let $\Omega_S(t)$ denote a region that encloses a fluid
system, that is a fixed material portion of the flow. In this case the velocity
of the surface $\partial\Omega_S(t)$ at any point is precisely the fluid
velocity at that point. Let $\boldsymbol{\eta}(\vec x, t)$ denote a (volume)
density describing some property of the fluid (like mass density, momentum
density, angular momentum density, etc.). The corresponding extensive property
for the system is then
\[
\vec{N}_S(t) = \int_{\Omega_S(t)} \boldsymbol{\eta}(\vec{x},t)\,d\x.
\]
The transport theorem states that
\begin{equation}\label{TTsystem}
\frac{d\vec{N}_S}{dt} = \int_{\Omega_S(t)} \frac{\partial \boldsymbol{\eta}}{\partial
t}\,d\x + \oint_{\partial \Omega_S(t)} \boldsymbol{\eta} \vec{u} \cdot \hat{n}\, dS.
\end{equation}
Here $\vec u$ denotes the fluid velocity, and $\hat n$ is the outward directed
normal to the boundary $\partial\Omega_S.$

Now, let $\Omega_{C}(t)$ be a control volume (a region of fluid defined for
convenience that does not necesarily move with the flow) with arbitrary
movement, and let $\vec{v}(\vec{x},t)$ denote the velocity of a
point at the surface $\partial\Omega_{C}(t).$ In this case we have
\begin{equation}\label{TTCV}
\frac{d}{dt} \int_{\Omega_{C}(t)} \boldsymbol{\eta}\,d\x =
\int_{\Omega_{C}(t)} \frac{\partial \boldsymbol{\eta}}{\partial t}\,d\x +
\oint_{\partial\Omega_{C}(t)} \boldsymbol{\eta} \vec{v} \cdot \hat{n}\, dS.
\end{equation}

Now, at a particular time of interest we choose a control volume $\Omega_C(t)$
which is concident with a system volume $\Omega_S(t)$, but not in general at
future times. That is $\Omega_C(t) = \Omega_S(t)$, but $\Omega_C(t') \neq
\Omega_S(t'),$ if $t' \neq t.$ Then we can eliminate the first term on the
right hand side in \eqref{TTsystem} by using \eqref{TTCV} which gives,
\begin{equation}\label{TT}
\frac{d\vec{N}_S}{dt} = \frac{d}{dt}\int_{\Omega_{C}(t)} \boldsymbol{\eta}\,d\x
+ \oint_{\partial\Omega_{C}(t)} \boldsymbol{\eta} (\vec{u} - \vec{v}) \cdot
\hat{n}\, dS.
\end{equation}
Notice that $\vec u - \vec v$ measures the fluid velocity at a boundary point
with respect to that boundary point.

We are interested in two particular cases. One of them is when $\vec{N}=\vec{P}$
is the total momentum contained in $\Omega_S(t),$ so that $\boldsymbol{\eta} =
\rho(\x,t)\u(\x,t)$.  The second case is when $\vec{N}=\vec{H}$ is the total
angular momentum, with respect to a reference point $\x_0$, so that 
$\boldsymbol{\eta} = \r(\x) \times \rho(\x,t)\u(\x,t),$ with $\r(\x)=\x-\x_0.$
The evaluation of equation \eqref{TTsystem} or \eqref{TT} for the momentum and angular 
momentum cases give us the total force and torque applied over the fluid system
contained in $\Omega_S(t) = \Omega_C(t)$. 

The first term on the right hand side in \eqref{TT} represents the
total variation of $\boldsymbol{\eta}$ contained in the control volume
$\Omega_C(t)$, while the second term in the right hand side is a surface
integral that amounts the $\boldsymbol{\eta}$ flowing out of the volume
$\Omega_C(t)$.

\subsection{Derivation of momentum exchange}

As explained in Section \ref{subsec:BoundaryConditions}, the boundary conditions
can be thought as an artificial flow inside $\Omega.$ This artificial  flow is
in turn decomposed into $Q$ artificial flows, one for each fundamental velocity
$\ci$ in the method. To explain the effect of these flows we refer back to the
figure \ref{fig_01}. Consider the boundary node $F$  and the direction $\ci$
pointing from $A$ to $F$. At every time step, the rol of the boundary condition
is to replace the value of $\hat f_i(\x_A, t)$ that would otherwise be provided
by a collision step, by a new value. Altogether, these replacements carried out
by the boundary condition are a way of introducing a certain amount of momentum
in the $i$ direction, at every time step. We derive ME by computing the amount
of momentum that the boundary condition introduces per unit time. In this way we
compute the force that each of these artifficial flows apply to the external
flow. The addition over all elementary directions $i$ accounts for the total
force the submerged body applies over the surrounding flow. By action-reaction
principle, the force that the surrounding flow applies over the the submerged
body is exactly the opposite.

We consider the system of particles associated to a lattice velocity $\ci$ that
at time $t$ is exactly inside $\Omega(t)$. At $t+\delta t$ this system moves by
an amount $\ci\delta t.$ We call ${\cal P}_{i,t}(t')$ the set of nodes
associated to this system of particles at time $t'$ and $\vec P_{i,t}(t')$
denotes its momentum at time $t'$. Finally we denote ${\cal
A}_t$ the set of lattice nodes $A$ inside $\Omega(t).$

In the following subsections we derive the force and torque that the flow
applies to the body through its surface. The cases of static and moving bodies
are treated.

\subsubsection{Force}\label{sssec:Force}

The amount of momentum the boundary conditions add per unit time to
the $i$-th system of particles is 
\begin{equation}\label{dPi}
\frac{d\vec P_{i,t}}{dt} = \frac{\vec P_{i,t}(t+\delta t) - \vec P_{i,t}(t)}{\delta
t} + {\cal O}(\delta t)
\end{equation}
where 
\begin{equation}\label{eq:approx_Momentum}
\vec{P}_{i,t}(t') = {\delta x}^D \sum_{A\in {\cal P}_{i,t}(t')} 
\ci f_{i}(\x_{A},t').
\end{equation} 
Neglecting ${\cal O}(\delta t)$ terms we have
\begin{multline}\label{eq:Force_general}
\frac{d\vec P_{i,t}}{dt} \simeq \frac{{\delta x}^D}{\delta t}\Bigl(
\sum_{A\in {\cal P}_{i,t}(t+\delta t)} \ci f_{i}(\x_{A},t+\delta t) \\
-\sum_{A\in {\cal P}_{i,t}(t)} \ci f_{i}(\x_{A},t) \Bigr).
\end{multline}

\subsubsection{Force on a static body}\label{sssec:Forces in static body}

\begin{center}
\begin{figure}[h]
\centerline{\includegraphics[scale=0.18]{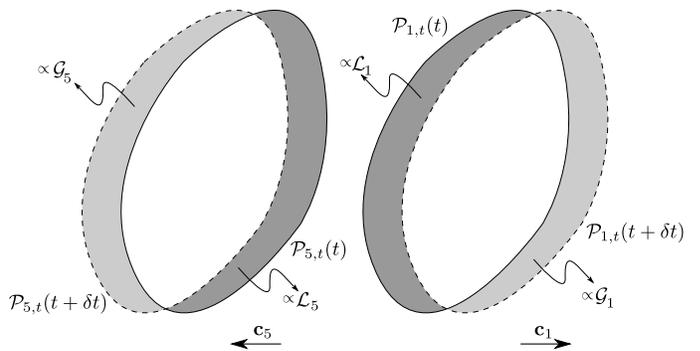}}
\caption{Schematic diagram of the areas occupied by 
${\cal P}_{i,t}(t)$ and ${\cal P}_{i,t}(t+\delta t)$ for $i=1,5$. 
The figure shows shaded areas proportional to the size of the sets 
$\Gi$ \emph{gained} and $\Li$ \emph{lost} nodes when 
${\cal P}_{i,t}(t)$ is displaced one lattice site in the $\vec{c}_5$ (left) and 
$\vec{c}_1$ (right) directions in the $D2Q9$ model.}
\label{fig_02}
\end{figure}
\end{center}

We assume first the case of a static body, so that $\Omega$ and the set ${\cal
A}$ are constant in time.  The first term in \eqref{eq:Force_general} can be
rewritten in terms of the sets $\Gi$ of {\em gained} and $\Li$ of {\em lost}
nodes as a consequence of the displacement of the system of particles from $t$
to $t+\delta t$. This displacement is exemplified in Figure \ref{fig_02} for the
$D2Q9$ model and the directions $i=1$ and $i=5.$

To simplify notation we define $g_i = \ci f_{i}(\x_{A},t+\delta t)$. The first
term in \eqref{eq:Force_general} becomes
\begin{equation}\label{eq:ForceTermAt_t+dt1}
\sum_{A\in {\cal P}_{i,t}(t+\delta t)} 
g_i = \sum_{A\in\Gi} g_i - \sum_{A\in\Li} g_i
 + \sum_{A\in {\cal P}_{i,t}(t)} g_i
\end{equation}
%
Inserting this into \eqref{eq:Force_general} and adding over the $Q$ systems
we get the LBM approximation to the force introduced by the boundary conditions.
\begin{multline}\label{constraint_force}
\vec{F}_{c}(t)\simeq {\delta x}^D \sum_{i=0}^{Q-1} \sum_{A\in {\cal P}_{i,t}(t)} \ci
\frac{f_{i}(\x_{A},t+\delta t) - f_{i}(\x_{A},t)}{\delta t} \\
+ \frac{{\delta x}^D}{\delta t} \sum_{i=0}^{Q-1} \Bigl(\sum_{A\in\Gi} g_i  
- \sum_{A\in\Li} g_i\Bigr)
\end{multline}
%
We want to compare this expression with the Reynolds transport theorem
\eqref{TT} applied to the artificial flow inside $\Omega(t)$. The force
introduced by the boundary conditions is the constraint force acting on the body
to keep it at a fixed position. The first term in the right hand side of
\eqref{constraint_force} is an LBM approximation of the volume term in
\eqref{TT}. The second term in \eqref{constraint_force} is composed of sums on
sets of nodes which are near neighbours of the boundary $\partial \Omega$.  This
second term is precisely the LBM approximation to the surface integral term in
\eqref{TT}.  As the interaction between the body and the surrounding fluid
occurs only through the body's surface, this second term in
\eqref{constraint_force} is the term we are interested in. By action-reaction
principle the flow force on the body is,
\begin{equation}\label{flow_force}
\vec{F}_{f}(t) \simeq \frac{{\delta x}^D}{\delta t} 
\sum_{i=0}^{Q-1} \Bigl(- \sum_{A\in\Gi} g_i + \sum_{A\in\Li}  g_i\Bigr)
\end{equation}
%
Notice that 
$A \in \Li$ if and only if there is a node $B \in {\cal
G}_{\bar i}$ such that $\x_A = \vec{x}_B + \ci\delta t.$ Therefore
\begin{multline*}
\vec{F}_{f}(t) \simeq \frac{{\delta x}^D}{\delta t} 
\sum_{i=0}^{Q-1} \Bigl( -\sum_{A\in\Gi} \ci f_{i}(\x_{A},t+\delta t) \\
+ \sum_{A\in{\cal G}_{\bar i}} \ci f_i(\x_{A} + \ci \delta t,t+\delta t)\Bigr)
\end{multline*}
Now, a sum over all sets ${\cal G}_{\bar i}$ can be written as a sum over all
sets $\Gi$, we obtain
\begin{multline}
\vec{F}_{f}(t) \simeq - \frac{{\delta x}^D}{\delta t} \sum_{i=0}^{Q-1}
\sum_{A\in\Gi} \ci \Bigl( f_{i}(\x_{A},t+\delta t) \\ 
+ f_{\bar i}(\x_{A}+\vec{c}_{\bar i}\delta t,t+\delta t)\Bigr).
\end{multline}
We notice that
\begin{equation}\label{eq:Equalities}
\begin{split}
f_{\bar i}(\x_{A}+\vec{c}_{\bar i}\delta t,t+\delta t) &= 
\hat f_{\bar i}(\x_A,t), \\
f_{i}(\x_{A},t+\delta t) &= \hat f_i(\x_A - \ci \delta t, t).
\end{split}
\end{equation}
The first identity is the streaming step from the outer nodes in a
direction that points into $\Omega$ (across the boundary). This values of $\hat
f_i$ are provided by the collision step. The second identity
is a streaming step from inner nodes in a direction pointing outwards (across the
boundary); these value of $\hat f_i$ are provided by the boundary condition.
The flow force on the sumberged body can then be written as
\begin{equation}\label{flow_force_2}
\vec{F}_{f}(t) \simeq - \frac{{\delta x}^D}{\delta t} \sum_{i=0}^{Q-1}
\sum_{A\in\Gi} \ci \bigl(\hat f_i(\x_A - \ci
\delta t, t) + \hat f_{\bar i}(\x_A, t)\bigr).
\end{equation}
To compare the equation \eqref{flow_force_2} with the equivalent ones in the
literature, care has to be taken as regards different definitions of the sets
$\Gi.$ Equation \eqref{flow_force_2} is precisely the expression that appears
extensively in the literature
\cite{Aidun:1998,Mei:2002_pre65-041203,Li:2004,Wen:2012} as
the momentum exchange method to evaluate forces in static bodies.

\subsubsection{Force on a moving body}\label{sssec:Forces in moving body}

For the case of a moving body we show two alternative derivations of the
flow force. In this way we recover the two main proposals that appeare in the
literature.

\begin{center}
\begin{figure}[t]
\includegraphics[scale=0.18]{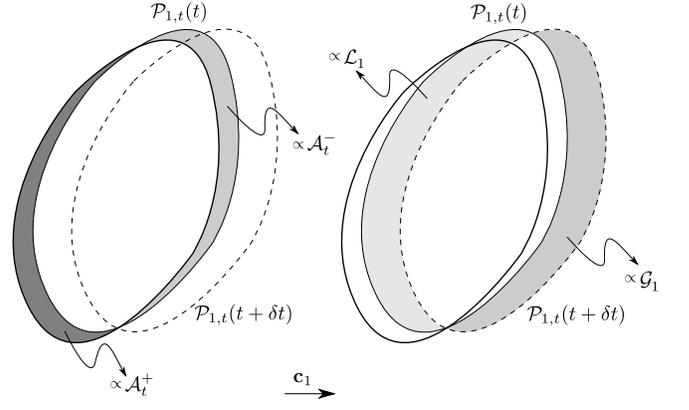}
\caption{Schematic diagram of the area occupied by the nodes 
${\cal P}_{1,t}(t)$ and ${\cal P}_{1,t}(t+\delta t)$.
The figure shows shaded areas proportional to the size of the lattice
nodes ${\cal A}^+_t$ and ${\cal A}^-_t$ (left), and ${\cal G}_1$ and ${\cal L}_1$
(right) as defined in the text.}\label{fig_03}
\end{figure}
\end{center}

When the submerged body is moving, the region $\Omega(t)$ and the set of lattice
nodes ${\cal A}_t$ are no longer constant. 
For some time steps, one can even expect the set of nodes ${\cal A}_{t+\delta t}$ 
to be the same as the set of nodes ${\cal A}_t.$ In any case it is useful define 
the sets of nodes ${\cal A}^+_t$ and ${\cal A}^-_t$ as
\[\begin{split}
A&\in{\cal A}^+_t, \quad \mbox{if} \quad A\in{\cal A}_{t+\delta t} \quad
\mbox{and} \quad A\notin{\cal A}_t,\\
A&\in{\cal A}^-_t, \quad \mbox{if} \quad A\in{\cal A}_t \quad
\mbox{and} \quad A\notin{\cal A}_{t+\delta t}.
\end{split}
\]
Figure \ref{fig_03} shows a scheme of a typical situation when the body moves.

The expression \eqref{eq:ForceTermAt_t+dt1} is still valid in this case. 
However, at time $t+\delta t$ we want to make reference to the body's new
position, so we rewrite the term that sums over ${\cal P}_{i,t}(t)$ as
\begin{equation}\label{eq:ForceTermChangingByMovement}
\sum_{A\in {\cal P}_{i,t}(t)} g_i =
\sum_{A\in{\cal{A}}_{t+\delta t}} g_i
+\sum_{A\in{\cal{A}}^{-}_{t}} g_i - 
\sum_{A\in{\cal{A}}^{+}_{t}} g_i
\end{equation}
%
We insert \eqref{eq:ForceTermChangingByMovement} into \eqref{eq:ForceTermAt_t+dt1}
and use the result into \eqref{eq:Force_general}. Then we add over $i$ to get an
approximation of the flow force acting on the body
$\Omega$
\begin{multline}\label{eq:ForceInMovement}
\vec{F}_{f}(t) \simeq \frac{{\delta x}^D}{\delta t}
\sum_{i=0}^{Q-1} \Bigl(
- \sum_{A\in\Gi} \ci \bigl(\hat f_i(\x_A - \ci \delta t, t) 
+ \hat f_{\bar i}(\x_A, t) \bigr) \\
- \sum_{A\in{\cal{A}}^{-}_{t}} \ci f_{i}(\x_{A},t+\delta t)
+ \sum_{A\in{\cal{A}}^{+}_{t}} \ci f_{i}(\x_{A},t+\delta t) \Bigr).
\end{multline}
Where again, as we are looking for the surface contributions to the force, we
dropped the volume contribution. Equation \eqref{eq:ForceInMovement} shows a
main term, which is the same as in the case of a static body, representing the
particle's exchange of momentum across the boundary, but now this term is
corrected by the last two terms which accounts for the momentum associated to
the nodes that enter or leave $\Omega(t)$ as a consequence of the body
movement. In this way we obtain terms similar to that proposed
by Aidun et. al. \cite{Aidun:1998} to evaluate the force on a moving body.  We
show that these terms are correct and necessary to obtain the complete
superficial contribution to the force when the body moves. Equation 
\eqref{eq:ForceInMovement} is then similar to that introduced in
\cite{Aidun:1998} and by Wen et. al. \cite{Wen:2012}, extensively 
used in the literature to evaluate the fluid force on moving bodies.

There is a minor difference between the expression \eqref{eq:ForceInMovement}
and those introduced in \cite{Aidun:1998} and \cite{Wen:2012}. In their cases,
the force at time $t$ considers the lattice nodes that enter and leave
$\Omega(t)$ between $t-\delta t$ and $t$ (i.e., backward in time). In our case,
\eqref{eq:ForceInMovement} requires to know the sets ${\cal{A}}^{+}$ and
${\cal{A}}^{-}$, that is the sets of nodes that enter and leave $\Omega$ between
$t$ and $t+\delta t$ (i.e., forward in time). The determination of the sets
${\cal{A}}^{+}$ and ${\cal{A}}^{-}$ is direct if the movement of the body is
given (predetermined) at all times, in this case \eqref{eq:ForceInMovement} is
an explicit expression. If, however, the motion of the body is to be computed
simultaneously with the flow, the equation \eqref{eq:ForceInMovement} becomes
implicit. In this last case it is convenient to use an approximation to
determine ${\cal{A}}^{+}$ and ${\cal{A}}^{-}$ so that the equation becomes
explicit.

In the numerical tests in section \ref{sec:numerical_tests}, we implement two
different approximations to find the sets ${\cal{A}}^{+}$ and ${\cal{A}}^{-}$.
Both approximations work well, giving no appreciable difference in the outcomes
of the benchmark tests. The first approximation is the procedure proposed in
\cite{Aidun:1998}. The second approximation is more complicated. It computes the
sets ${\cal{A}}^{+}$ and ${\cal{A}}^{-}$ by approximating the region
$\Omega(t+\delta t)$ as if it was moving with the speed computed at the previous
time step. With this information the flow force can be computed at time $t$ and
then the correct displacement of $\Omega$ from $t$ to $t+\delta t$ 
recomputed. Though computationally more expenssive, as two displacements of
$\Omega$ are computed at each time step, this second approximation is more
precise than the first one and may be worth using it in some situations.

Notice that the variables associated to the lattice nodes belonging to ${\cal
A}^-$ do not have values assigned at time $t$ since these nodes enter the fluid
region between $t$ and $t+\delta t$. These values are needed in order to compute
the time step from $t$ to $t+\delta t$. As mentioned previously, various rules
to ``initialize'' these variables are proposed in the literature. In our
simulations we implement the proposals given in \cite{Lallemand:2003} and
\cite{Aidun:1998}. Also we implement a method that sets the mentioned variables
by using the equilibrium distribution function, where the macroscopic variables
are set as an average of the values at the nearest neighbor fluid nodes. The
evaluation of the force by \eqref{eq:ForceInMovement} we present in Section
\ref{sec:numerical_tests} show a short time scale noise. The use of the first
two methods mentioned before to initialize the nodes that enter the fluid region
present lower noise level.

The main sources of ``noise'' in the force evaluation using
\eqref{eq:ForceInMovement} are the impulsive nature
of the additional terms related to ${\cal{A}}^{+}$ and ${\cal{A}}^{-}$. This
noise have been observed before. In \cite{Wen:2014, Krithivasan:2014} the
authors  show some alternative methods to avoid this undesirable effect.  As the
time derivative of the momentum $\vec P_{i,t}$ is independent of the inertial reference
frame, we can recover these methods by repeating the derivation we did before by
choosing, for each lattice node $\x_A$ and direction $i,$ a convenient reference
frame. For those nodes which are close to the boundary and for each direction
$i$ pointing to the boundary we express the momentum in the reference frame in
which the velocity $\vec v_{Ai}$  of the intersection point of the boundary with
the lattice link joining $\x_A - \ci\delta t$ with $\x_A$ is zero. The interior
nodes that are far from the boundary contribute only to a volume term in the
force, this volume term is dropped and therefore the reference frame is
unimportant. The result obtained in this way is an LBM discretization of the surface
term in the right hand side of
\eqref{TT}
\begin{multline}\label{eq:ForceInMovementMovingFrame}
\vec{F}_{f}(t) \simeq \frac{{\delta x}^D}{\delta t}
\sum_{i=0}^{Q-1} \Bigl(
- \sum_{A\in\Gi} (\ci - \vec{v}_{Ai}) \hat f_i(\x_A - \ci \delta t, t) \\ 
- (\vec c_{\bar i} - \vec{v}_{Ai}) \hat f_{\bar i}(\x_A, t)\Bigr)\\ 
- \sum_{A\in{\cal{A}}^{-}_{t}} \sum_{i=0}^{Q-1} (\ci - \vec{v}_{Ai})
f_{i}(\x_{A},t+\delta t) \\
+ \sum_{A\in{\cal{A}}^{+}_{t}} \sum_{i=0}^{Q-1} (\ci - \vec{v}_{Ai})
f_{i}(\x_{A},t+\delta t).
\end{multline}
The last two terms in the right hand side of \eqref{eq:ForceInMovementMovingFrame}
are negligible since both $\sum_{i=0}^{Q-1}\ci f_i$ and
$\sum_{i=0}^{Q-1} \vec v_{Ai} f_i$ represent close approximations to $\rho \vec
u$ at the boundary points. 

Either expressions \eqref{eq:ForceInMovement} and
\eqref{eq:ForceInMovementMovingFrame} are correct expressions; they constitute
different approximations of the flow force. The later has some advantages
though. First, it is computationally more efficient, since it is not necessary
to determine the sets ${\cal A}^+_t$ and ${\cal A}^-_t$. As a result the method
is always explicit and it presents a notorius noise decrease in force
evaluation as shown in \cite{Wen:2014}.

\subsubsection{Torque}\label{sssec:Torque}

The derivation of the torque acting on the submerged body is analogous to that
of the force. The angular momentum per unit time introduced by the $i$-th
artificial flow is
\begin{equation}\label{eq:dHdt}
\frac{d\vec{H}_{i,t}}{dt} = 
\frac{\vec H_{i,t}(t+\delta t) - \vec H_{i,t}(t)}{\delta t} + {\cal O}(\delta t) 
\end{equation}
where
\begin{equation}\label{eq:H}
\vec{H}_{i,t}(t') = {\delta x}^D \sum_{A\in {\cal P}_{i,t}(t')} 
\r(\x_{A}) \times \ci f_{i}(\x_{A},t'),
\end{equation}
with $\vec r(\x_A) = \x_A - \x_0,$ $\vec{H}_{i,t}(t')$ is the angular momentum
of the particle system at time $t'$ with respect to a fixed point $\x_0.$
Neglecting ${\cal O}(\delta t)$ terms in equation \eqref{eq:dHdt} we have
\begin{multline}\label{eq:Torque_general}
\frac{d\vec H_{i,t}}{dt} \simeq 
\frac{{\delta x}^D}{\delta t}\Bigl(
\sum_{A\in{\cal P}_{i,t}(t+\delta t)} \r(\x_{A}) \times \ci f_{i}(\x_{A},t+\delta t) \\
- \sum_{A\in{\cal P}_{i,t}(t)} \r(\x_{A}) \times \ci f_{i}(\x_{A},t) \Bigr).
\end{multline}

As we have done in section \ref{sssec:Force}, we treat the case of a static body
first and then extend the proposal to the case of a moving body.

\subsubsection{Torque on static body}\label{sssec:Torque in static body}

Using the lattice nodes sets $\Gi$ and $\Li$ 
(shown in Figure \ref{fig_02}) to rewrite the first term in \eqref{eq:Torque_general},
and denoting  $h_i=\r(\x_{A}) \times \ci f_{i}(\x_{A},t+\delta t)$ for simplicity,  
we have
\begin{equation}\label{eq:TorqueTerm_t+dt}
\sum_{A\in{\cal P}_{i,t}(t+\delta t)} h_i = 
 \sum_{A\in{\cal P}_{i,t}(t)} h_i + \sum_{A\in\Gi} h_i - \sum_{A\in\Li} h_i 
 \end{equation}
%
Inserting this into \eqref{eq:Torque_general} and adding over the $Q$ systems
we get an approximation to the constraint torque acting on $\Omega,$ 
\begin{multline}\label{constraint_torque}
\vec{T}_{c}(t)\simeq \\ {\delta x}^D \sum_{i=0}^{Q-1} 
\sum_{A\in{\cal P}_{i,t}(t)} \r(\x_{A}) \times \ci \frac{f_{i}(\x_{A},t+\delta
t) - f_{i}(\x_{A},t)}{\delta t} \\
+ \frac{{\delta x}^D}{\delta t} \sum_{i=0}^{Q-1} \Bigl(
\sum_{A\in\Gi} h_i - \sum_{A\in\Li} h_i
\Bigr)
\end{multline}
%
As in the force case, we can compare this expression with the Reynolds Transport
theorem, then keeping just the approximation of the surface term in
\eqref{TT}, we get an expression for the torque that the flow applies on
the body,
\begin{equation}\label{eq:flow_torque}
\vec{T}_{f}(t)\simeq \frac{{\delta x}^D}{\delta t} \sum_{i=0}^{Q-1} \Bigl(
- \sum_{A\in\Gi} h_i 
+\sum_{A\in\Li} h_i
\Bigr)
\end{equation}
%
Recalling the relation between $\x_A \in \Li$ and 
$\vec{x}_B \in {\cal G}^{\bar i}$ ($\x_A = \vec{x}_B + \ci\delta t$), and using
\eqref{eq:Equalities}
\begin{multline}\label{eq:flow_torque_2}
\vec{T}_{f}\simeq -\frac{{\delta x}^D}{\delta t} \sum_{i=0}^{Q-1} 
\sum_{A\in\Gi} \Bigl(  \r(\x_{A}) \times \ci \bigl( 
\hat{f}_{i}(\x_{A}-\ci \delta t,t) \\
+ \hat{f}_{\bar{i}}(\x_{A},t)\bigr) \Bigr)
\end{multline}
This equation is the expression that appears in the literature
\cite{Aidun:1998,Mei:2002_pre65-041203,Li:2004,Wen:2012}
extensively as the momentum exchange method to evaluate torque on static bodies.

\subsubsection{Torque on a moving body}

For a moving body we follow a procedure and reasoning analogous to that of
section \ref{sssec:Forces in moving body}. We rewrite the first term on the
right hand side of \eqref{eq:TorqueTerm_t+dt} to get the correct surface contribution
when the surface moves,
\begin{equation} \label{eq:TorqueTermChangingByMovement}
\sum_{A\in{\cal P}_{i,t}(t)} h_i = \sum_{A\in{\cal A}_{t+\delta t}} h_i  
+\sum_{A\in{\cal{A}}^{-}_{t}} h_i - \sum_{A\in{\cal{A}}^{+}_{t}} h_i 
\end{equation}
%
We replace \eqref{eq:TorqueTermChangingByMovement} in 
\eqref{eq:TorqueTerm_t+dt}, then from equation \eqref{eq:Torque_general} and adding 
over the $Q$ systems we obtain an approximation of the constraint torque
acting on the body at time $t.$ Thus the flow torque on a
moving body turns out to be
\begin{multline}\label{eq:TorqueInMovement}
\vec{T}_{f}(t) \simeq -\frac{{\delta x}^D}{\delta t}
\sum_{i=0}^{Q-1} \Bigl(\sum_{A\in\Gi} \r(\x_{A}) \times \ci \bigl(\hat
f_i(\x_A - \ci \delta t, t) \\
+ \hat f_{\bar i}(\x_A, t)\bigr) + \sum_{A\in{\cal{A}}^{-}_{t}} 
\r(\x_{A}) \times \ci f_{i}(\x_{A},t+\delta t) \\
- \sum_{A\in{\cal{A}}^{+}_{t}} \r(\x_{A}) \times \ci f_{i}(\x_{A},t+\delta t) \Bigr).
\end{multline}
Where we have used the relation of sets $\Gi$ and $\Li$, and the
equalities \eqref{eq:Equalities}.

The equation \eqref{eq:TorqueInMovement} has two distinct contribution to the
flow torque on $\Omega(t)$. The first one, is the contribution due to the exchange 
of momentum across the boundary as a consequence of the displacement of the
particle system from $t$ to $t+\delta t$. 
The second one, is the contribution to the torque by the lattice nodes that enter 
and leave $\Omega(t)$ as a consequence of its displacement to $\Omega(t+\delta t)$.
These are impulsive terms as we have noted in section \ref{sssec:Forces in moving body}.

Expression \eqref{eq:TorqueInMovement} is similar to the one presented in the
literature to evaluate the flow torque on moving bodies. This expression
naturally introduces the  ad-hoc correction terms first presented in
\cite{Aidun:1998} and used in \cite{Wen:2012}.
  
As with the force, a difference between our proposal and those in the literature
is the time at which the sets of lattice nodes ${\cal{A}}^{+}_t$ and
${\cal{A}}^{-}_t$ are evaluated. To avoid implicit expressions when the body
movement is not predefined, we use some approximation methods, presented in
section \ref{sssec:Forces in moving body}, to approach ${\cal{A}}^{+}_t$ and
${\cal{A}}^{-}_t$. 

As one could expect, some short time scale noise in the torque computation
appears as a consequence of the lattice nodes that enter and leave the fluid
domain as the body moves. 

As with the force derivation, we also obtain an alternative derivation for the
torque by considering the time derivatives of the angular momentum in different
reference frames for each particle. The reference frames to compute the torque
on the boundary nodes are chosen as in the derivation of
\eqref{eq:ForceInMovementMovingFrame}. The expression we get for the flow torque
on the body is
\begin{multline}\label{eq:TorqueInMovementMovingFrame}
\vec{T}_{f}(t) \simeq\\
-\frac{{\delta x}^D}{\delta t}
\sum_{i=0}^{Q-1} \Bigl(\sum_{A\in\Gi} \r(\x_{A}) \times (\ci - \vec v_{A,i}) \bigl(\hat
f_i(\x_A - \ci \delta t, t) \\
- \r(\x_A) \times (\vec c_{\bar i} - \vec v_{A,i}) \hat f_{\bar i}(\x_A,
t)\bigr) +\\
\sum_{A\in{\cal{A}}^{-}_{t}} \r(\x_{A}) \times (\ci - \vec v_{A,i})
f_{i}(\x_{A},t+\delta t) \\
- \sum_{A\in{\cal{A}}^{+}_{t}} \r(\x_{A}) \times (\ci - \vec v_{A,i})
f_{i}(\x_{A},t+\delta t) \Bigr).
\end{multline}
As with the force, the last two terms are negligible. Dropping these terms, the
expression becomes explicit and present lower noise level in the torque
evaluation.

\section{Numerical Tests}\label{sec:numerical_tests}

In this section we compare the results obtained with the expressions derived in section
\ref{sec:momentum_exchange} to compute the force and torque acting on a
submerged body. To this end we perform two benchmark tests on well known
problems that have been tested and benchmarked widely with others computational
fluid dynamics methods, such as finite element method and finite difference
methods.

We are interested in analyzing the dynamics of single bodies sedimenting
along a vertical channel filled with a Newtonian fluid. The bodies are either
circular or elliptic discs. The accuracy in the determination of the force
and torque acting on the falling body directly affects the body's movement. If
the force and torque are computed correctly, the displacement and rotation of
the bodies along the domain should be in agreement with data presented in the
literature \cite{Li:2004,Wen:2012,Feng:1994,Xia:2009}.

To solve the flow we use a \emph{D2Q9} lattice scheme and SRT with $\tau=0.6$.
The fluid density and the kinematic viscosity are set to $\rho_f=1000\,$kg/m$^3$
and $\nu=1\times10^{-6}\,$m$^2$/s respectively. The fluid is initially at rest
and has zero velocity at the horizontal and vertical boundaries at all times. We
implement these boundary conditions with the method presented in
\cite{Zou:1997}. The acceleration of gravity acting on the body is $g=9.81\,$m/s$^2$
downwards.

The motion of each body is determined by integrating Newton's equation of
motion, where the force is given by the fluid flow force, weight and buoyancy
force and the torque is given by the flow torque. To integrate in time we use
Euler Forward numerical scheme, which is first order accurate as the LBM method
itself. We have also implemented two step (Adams-Bashforth) integration in time
and noticed no appreciable difference in the results.

\vspace{1cm}

\subsection{Sedimentation of a circular disc}\label{subsec:SedimentationDisc}

\begin{center}
\begin{figure}[h]
\includegraphics[scale=0.18]{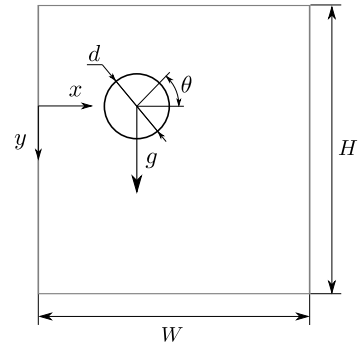}
\caption{An schematic diagram of the sedimentation disc problem.}
\label{fig_04}
\end{figure}
\end{center}

In this benchmark test we analyze the dynamics of a single two-dimensional disc
sedimenting along a vertical channel, shown schematically in Figure
\ref{fig_04}. We test the dynamics of the disc for two
density relations $r_\rho=\rho_b/\rho_f$, with $\rho_b$ and $\rho_f$
the densities of the body (disc) and the fluid respectively.
The dimensions of the vertical channel are $W=4d$ and $H=8W$; the disc diameter
is $d=1\times10^{-3}\unit{m}$. The disc center is initially placed at
$(x,y)=(7.6\times10^{-4},0)\unit{m}$ with the coordinate origin at
$2.5\times10^{-2}\unit{m}$ from the bottom of the channel and placed as shown
in Figure \ref{fig_04}.   We discretized the
computational domain with $n_x \times n_y = 135 \times 1073$ lattice points.

\begin{center}
\begin{figure}[ht]	
\includegraphics{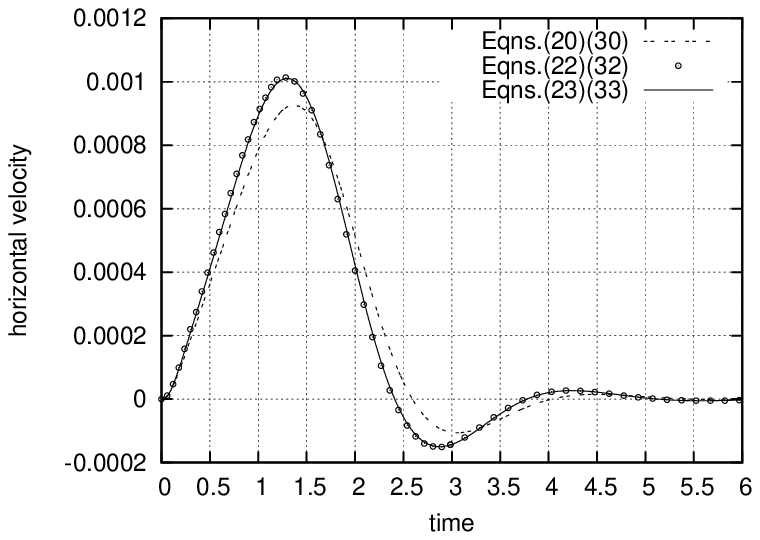}
\includegraphics{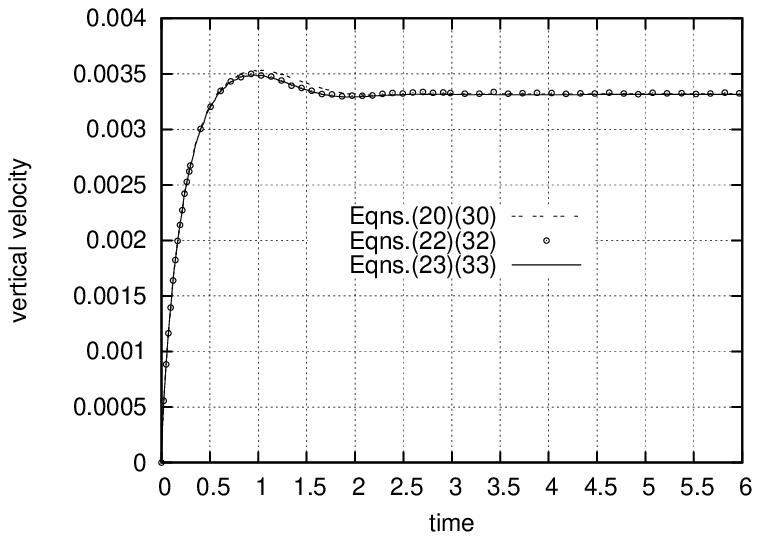}\\
\includegraphics{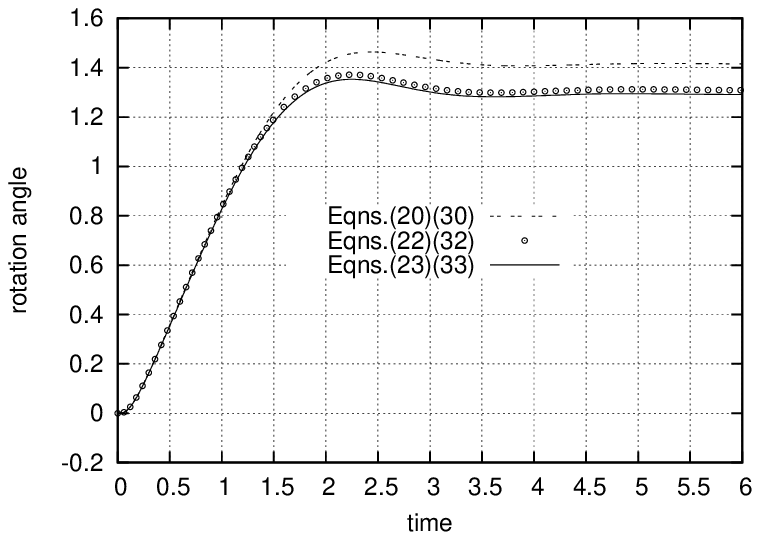}
\includegraphics{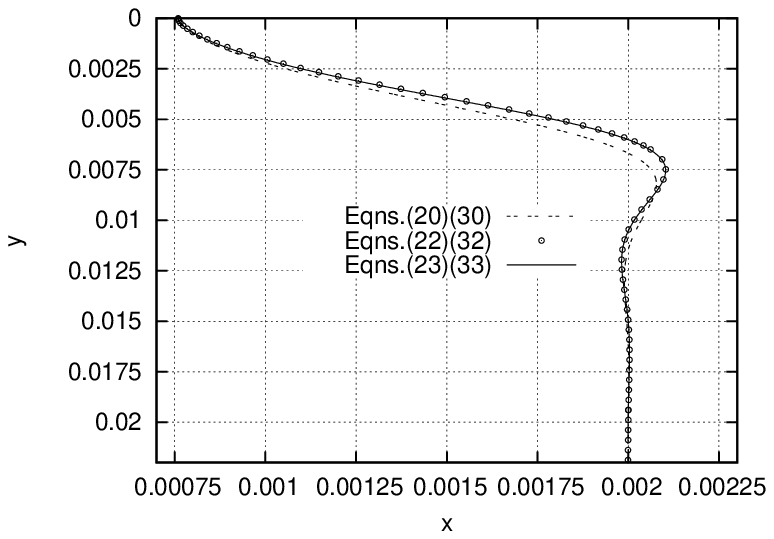}
\caption{Results obtained for the sedimenting disc of Figure
\ref{fig_04} for $r_\rho=1.01$. All magnitudes are
expressed in the international system of
units.}\label{fig_05}
\end{figure}
\end{center}

\begin{center}
\begin{figure}[ht]	
\includegraphics{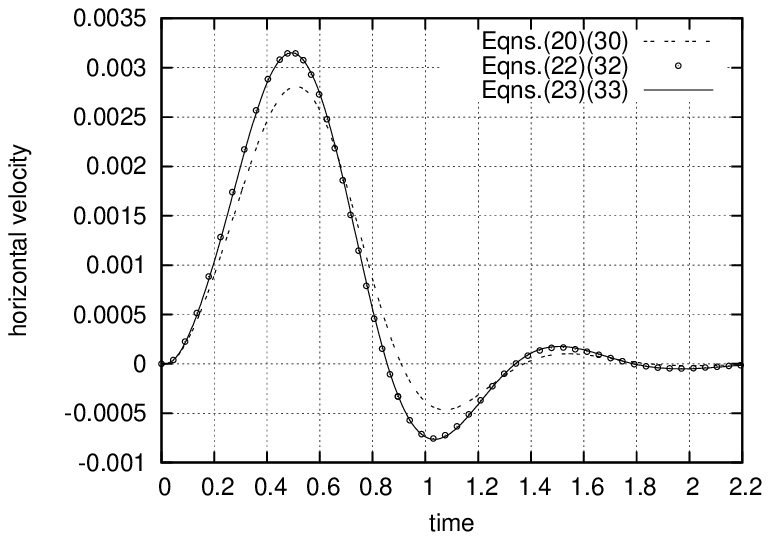}
\includegraphics{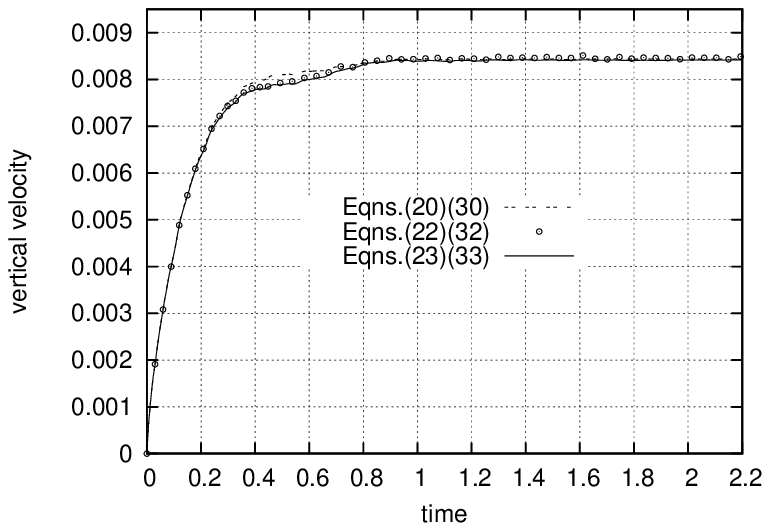}\\
\includegraphics{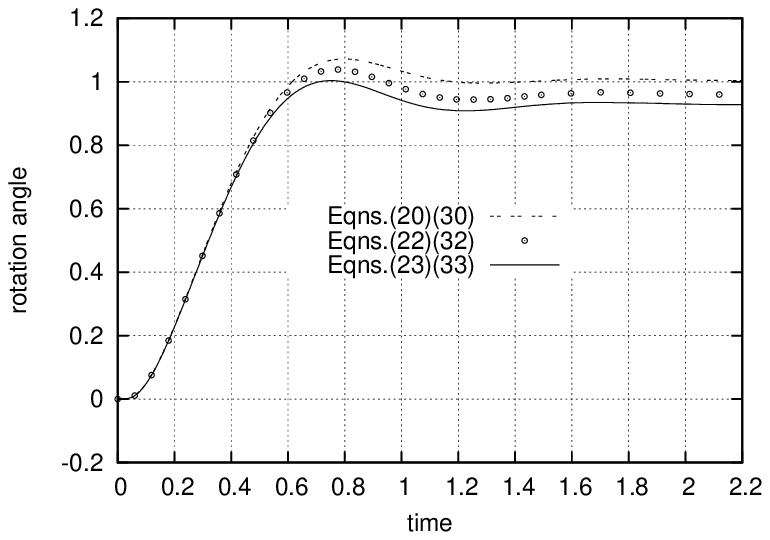}
\includegraphics{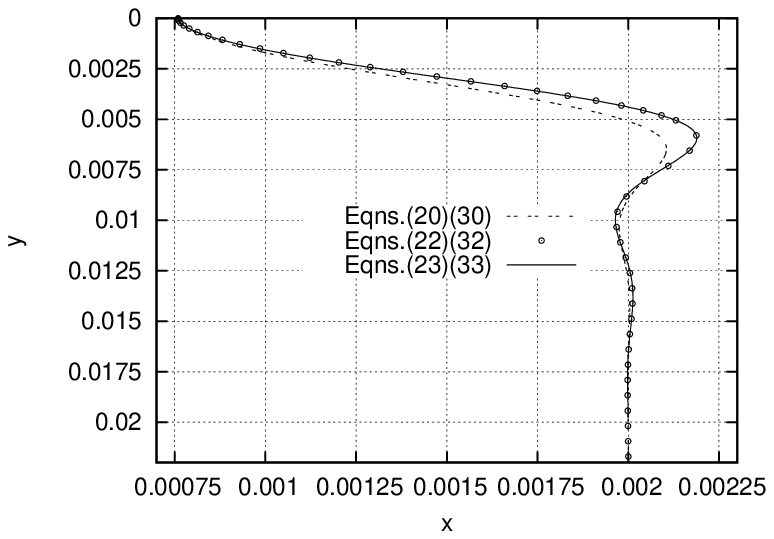}
\caption{Results obtained for the sedimenting disc of Figure
\ref{fig_04} for $r_\rho=1.03$. All magnitudes are
expressed in the international system of
units.}\label{fig_06}
\end{figure}
\end{center}

We test the performance of the method for two density ratios $r_\rho=1.01,$ and
$1.03.$ In Figures \ref{fig_05} and
\ref{fig_06} we show the horizontal and
vertical velocities and the trajectory of the center of the disc and the
rotation angle of the disc as functions of time, for $r_\rho=1.01$  and $r_\rho=1.03$.

When the disc is released from the initial position at $t=0$, it starts moving
and rotating along the channel. As one can see in the figures \ref{fig_05} and
\ref{fig_06}, the movement of
the disc can be divided into two regimes: A transient and a stationary regime. 

We compare results we obtained using a classical ME
\eqref{flow_force_2},\eqref{eq:flow_torque_2} and the corrected methods given by
\eqref{eq:ForceInMovement}, \eqref{eq:TorqueInMovement} and
\eqref{eq:ForceInMovementMovingFrame}, \eqref{eq:TorqueInMovementMovingFrame}.
These results, particularly those obtained with the corrected methods are in
good agreement with tests presented in \cite{Li:2004} (obtained using LBM with
SI), \cite{Wen:2012} (obtained using LBM with an expression similar to
\eqref{eq:ForceInMovement}, \eqref{eq:TorqueInMovement}) and \cite{Feng:1994}
(obtained using FEM). We observe visible discrepancies between the classical and
the corrected methods for the horizontal velocity and position. The major
discrepancy shows in the transient regime; no significant discrepancies can be
seen in the stationary regime.  Similar observations have been made by Wen et.
al.  \cite{Wen:2012} and Li et.  al \cite{Li:2004}.
\onecolumngrid
\twocolumngrid
\onecolumngrid
\twocolumngrid

\subsection{Sedimentation of an elliptic disc}\label{subsec:SedimentationEllipse}

In this section we present a benchmark test, similar to the previous one, where
the circular disc is replaced with an elliptical disc, also sedimenting in a
vertical channel filled with Newtonian fluid.  This test is also widely analyzed
in the literature. We study a problem as the one presented by Xia et. al.
\cite{Xia:2009}, where the authors use LBM and SI to obtain the forces
on the body.

\begin{figure}[ht]
\begin{center}
\includegraphics[scale=0.18]{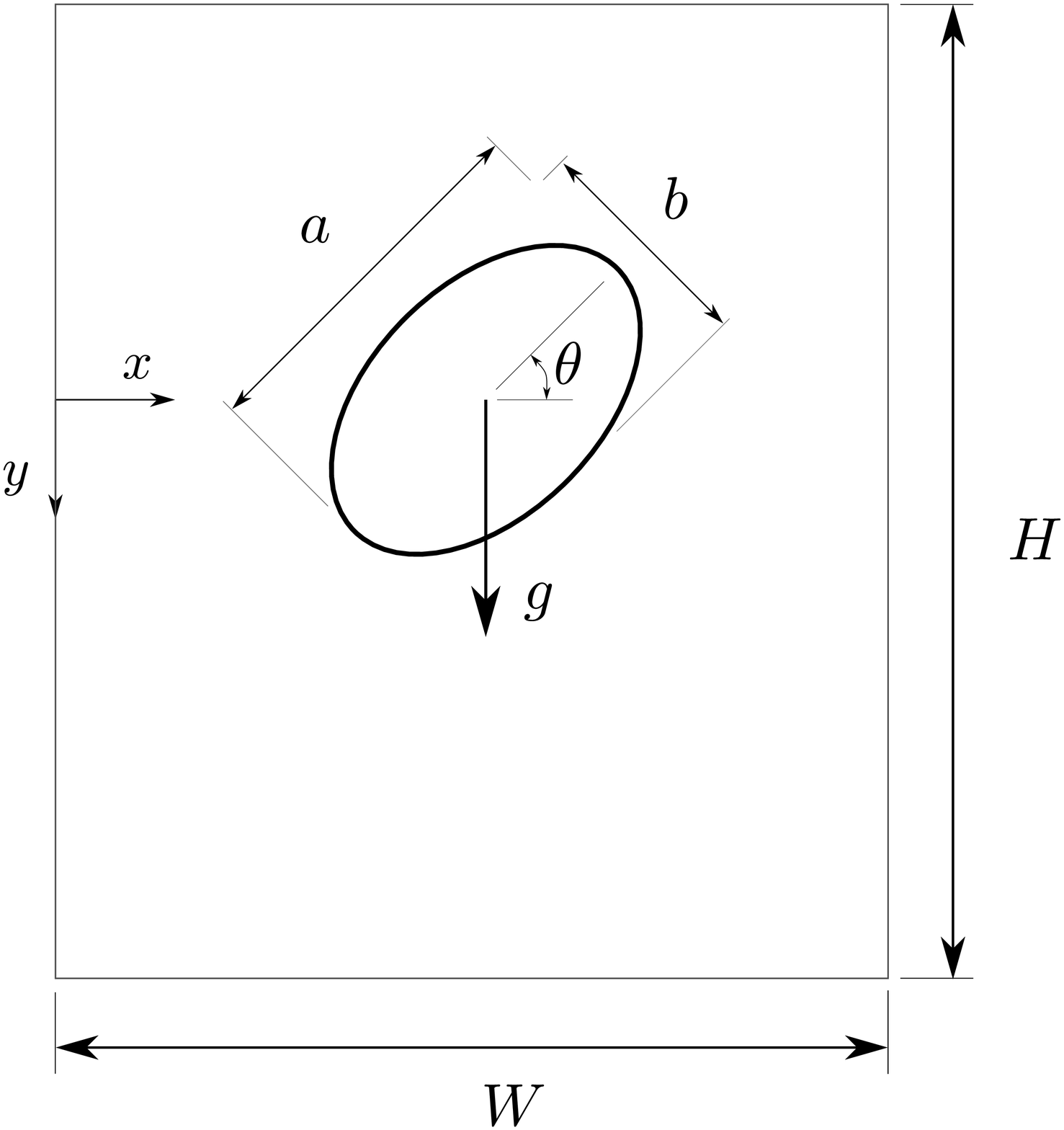}
\caption{An schematic diagram of the two-dimensional elliptical particle
sedimenting in a vertical channel.} \label{fig_07}
\end{center}
\end{figure}

We show in Figure \ref{fig_07} a schematic diagram of
the problem. We define three dimensionless parameters that characterize the
problem. These parameters are the aspect ratio $\alpha=a/b$, with $a$ and $b$
the major and minor axes of the ellipse respectively, the blockage ratio
$\beta=W/a$, with $W$ the width of the vertical channel, and the density ratio
$r_\rho$ as defined in Section \ref{subsec:SedimentationDisc}.

An exhaustive analysis of this sedimentation problem was carried out by Xia et.
al. \cite{Xia:2009}. They studied the influence on the dynamics of the density
ratio, the aspect ratio, and the channel blockage ratio. For simplicity we
analyze this problem with a fix blockage ratio, chosen so that we don't need to
consider the wall-particle interaction.  Our interest is to test the method
proposed in the present work, not to give a complete description of the
sedimentation problem.  We carry out simulations with a fixed geometrical
configuration.

In our tests we use major axis $a=10^{-3}$m, aspect ratio $\alpha=2$ and
blockage ratio $\beta=4.0$. The properties of the fluid are the same used in
Section \ref{subsec:SedimentationDisc}. Initially, the fluid is at rest, the
center of the ellipse is placed at $(x,y)=(0.5W,0)$m. The coordinate origin at
$4.8\times10^{-2}$m from the bottom of the vertical channel.  To break the
symmetry of the problem, we choose an initial angular position
$\theta_0=\frac{\pi}{4}$.  We set, following \cite{Xia:2009}, a height $H=50a$
and a width $W=4a$. The domain is discretized in a lattice with $n_x\times
n_y = 135\times 1676$ points and density ratio is $r_\rho = 1.10.$
 
\begin{center}
\begin{figure}[t]	
\includegraphics{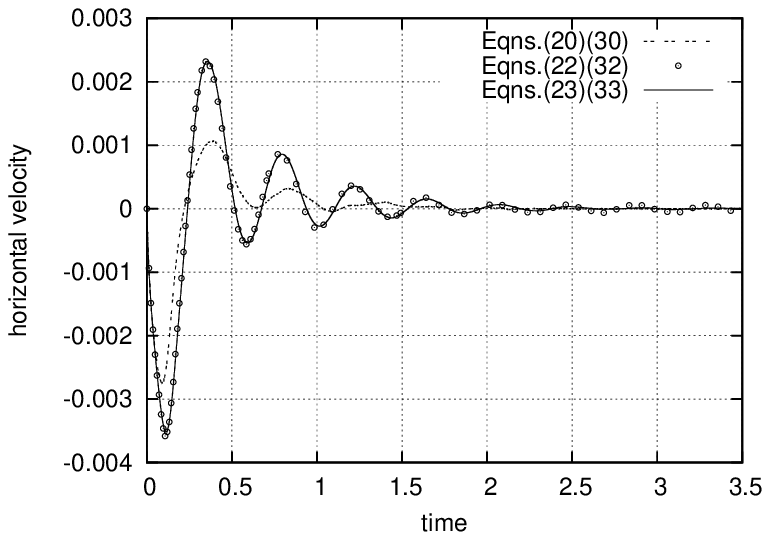}
\includegraphics{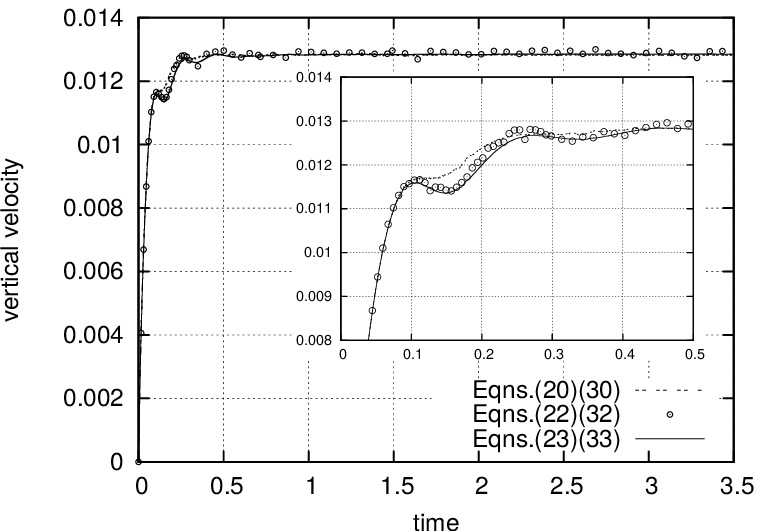}\\
\includegraphics{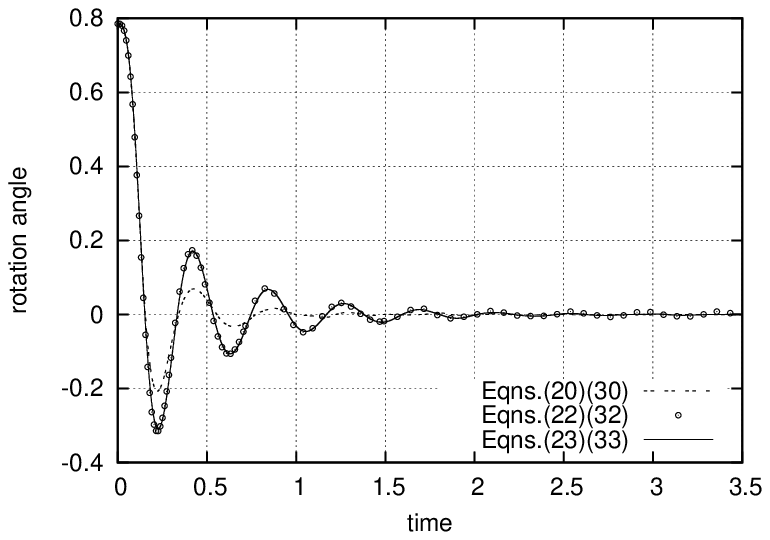}
\includegraphics{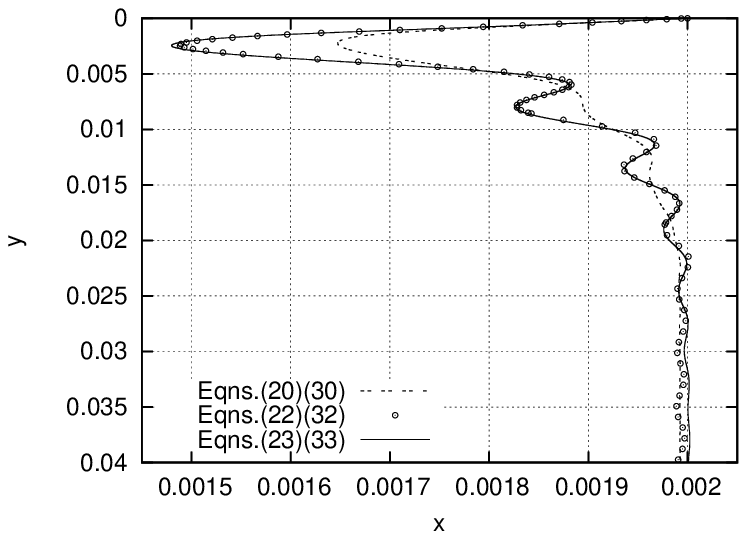}
\caption{Results for the sedimenting elliptical disc of Figure
\ref{fig_07} using $r_\rho=1.10.$ All magnitudes are expressed in the
international system of units.}\label{fig_08}
\end{figure}
\end{center}

In the Figure \ref{fig_08} we show the dynamical variables given as a function
of time and the complete trajectory of the ellipse computed using a classical ME
\eqref{flow_force_2},\eqref{eq:flow_torque_2} and the corrected methods given by
\eqref{eq:ForceInMovement}, \eqref{eq:TorqueInMovement} and
\eqref{eq:ForceInMovementMovingFrame}, \eqref{eq:TorqueInMovementMovingFrame}.
Our results using the corrected methods are in good agreement with the results
of  Xia et. al.  \cite{Xia:2009}. It is clear from Figure \ref{fig_08}, that
there exists an important difference, in the transient regime, and a minor
difference in the final horizontal position between the corrected and uncorrected
methods.

\section{Conclusion and discussion}\label{sec:comments}

In this work we have presented a new derivation of the momentum exchange method
to compute the flow force and torque acting on a submerged body. The expressions
we obtain, for the case of static bodies, are coincident with those presented in
\cite{Mei:2002_pre65-041203}. From our derivation we see that the expressions
derived for the flow force and torque on static bodies are not appropriate to
treat moving bodies. Moreover, we derive two of the proposals apeearing in the
literature to compute flow force and torque on moving bodies as particular
cases. These last two alternatives to compute the force and torque are correct
but different approximations to the same problem. The one consisting in
\eqref{eq:ForceInMovementMovingFrame} and \eqref{eq:TorqueInMovementMovingFrame}
results in less noisy force and torque computations and is also more efficient
from the computational point of view.

Our method of deriving momentum exchange does not use a particular treatment of
the boundary conditions on the body surface and  can be applied with several of
the various methods proposed in the literature.

In the last part of the paper we have tested the corrected momentum exchange
expressions we obtained by simulating two problems which are well know in the
literature, a sedimenting circular disc and a sedimenting elliptic. Our results
clearly show the difference, for the case of moving bodies, between the results
of the corrected momentum exchange methods as compared to those given by
equations \eqref{flow_force_2} and \eqref{eq:flow_torque_2}. These results are
in good agreement with those obtained by other authors using similar and
different computational fluid dynamic methods such as finite element methods.

\section*{Acknowledgments}
We want to thank Carlos Sacco and Ezequiel Malamud for useful discussions. J. P.
Giovacchini is a fellowship holder of CONICET (Argentina).  This work was
supported in part by grants 05-B454 of SECyT, UNC and PIDDEF 35-12 (Ministry of
Defense, Argentina). We want to thank the corrections and suggestions made by
the referees of the first manuscript that helped us improve our work.

\onecolumngrid
\twocolumngrid


\end{document}